# Measuring Urban Sprawl Based on Massive Street Nodes and the Novel Concept of Natural Cities

Tao Jia[(1)] and Bin Jiang[(2)]

[(1)]Future Position X, Box 975, 801 33 Gävle, Sweden
Email: jiatao83@163.com

[(2)]Department of Technology and Built Environment, Division of Geomatics
University of Gävle, 801 76 Gävle, Sweden
Email: bin.jiang@hig.se

**Abstract**
In this paper, we develop a novel approach to measuring urban sprawl based on street nodes and naturally defined urban boundaries, both extracted from massive volunteered geographic information OpenStreetMap databases through some data-intensive computing processes. The street nodes are defined as street intersections and ends, while the naturally defined urban boundaries constitute what we call natural cities. We find that the street nodes are significantly correlated with population of cities. Based on this finding, we set street nodes as a proxy of population to measure urban sprawl. We further find that street nodes bear a significant linear relationship with city areal extents. In the plot with the x axis representing city areal extents, and the y axis street nodes, sprawling cities are located below the regression line. We verified the approach using urban areas and population from the US census, and then applied the approach to three European countries: France, Germany, and the United Kingdom for the categorization of natural cities into three classes: sprawling, compact, and normal. This categorization sets a uniform standard for cross comparing sprawling levels across an entire country.

## 1. Introduction

Urban sprawl is generally characterized to be low density, auto-dependent land development taking place at the periphery of urban areas, and thus transfers a large amount of open space into low density suburban land. The loss of open space is not just a problem of rural land, but also has significant environmental and economic impacts (Cieslewicz, 2002, Burchell et al. 2005). Considerable effort has been devoted to the study of categorization of land use patterns and measurement of urban sprawl in order to provide guidance for spatial planning and policy making (Ewing et al. 2002, Cutsinger and Galster 2006, Weitz 1999, Fulton et al. 2001, Torrens 2008). Some of the studies took various variables such as housing mix, accessibility and measures of urban centers in the measuring process. To curb urban sprawl, we first need to effectively measure it to better understand how and where it has occurred (Burchfield et al. 2006). Existing studies as cited here mainly rely on population data and legally or administratively determined urban boundaries for measuring urban sprawl, although satellite imagery has been used to automatically detect urban boundaries (e.g., Sutton 2003, Sudhira et al. 2004, Ji et al. 2006). The population data are often available at an aggregate level, e.g., census tracts, designated places, or population centers; the population data acquired at an individual level, e.g., census blocks, are not freely available. The aggregated population data do not reflect real demographic facts. On the other hand, the legally or administratively determined urban boundaries are criticized for being subjective or even arbitrary. There is little consensus reached as to where the city boundary is, and how it is defined or delineated. In addition, city boundaries change over time, and the boundary has a major effect on drawing a conclusion on whether or not a city is sprawling. A fat boundary means that a city is sprawling, while a thin boundary could change the conclusion in the opposite direction.

We believe that measuring urban sprawl must be based on geospatial data at an individual level and with little ambiguity. In this respect, effectively defined city boundaries are very important for measuring urban sprawl, and they set a uniform standard for cross comparison. More than that, effectively defined city boundaries are of value to many urban studies for understanding the underlying structure and dynamics which concern economists, geographers, and even physicists. In this paper, we introduce a novel approach to measuring urban sprawl using naturally defined natural cities and street nodes that can be automatically derived from large openstreetmap (OSM) databases (Haklay and Weber 2008). Based on individuals' voluntarily contributions, and using free data such as GPS traces and uncopyrighted satellite imagery, the OSM community has generated over a few hundred of gigabytes of geospatial data, constituting one successful example of volunteered geographic information, supported by Web 2.0 technologies (Goodchild 2007, Sui 2008). We extracted a large amount of street nodes and street blocks from the massive OSM data in order to further derive individual natural cities through some clustering processes (*c.f.*, Section 2 for more details). We relied on the natural cities and street nodes for measuring urban sprawl, and found that the results are consistent with those reported in the literature.

The contributions of this paper are three-fold. First, a major contribution of this paper lies in the categorization of cities into sprawling or compact cities in a uniform standard for cross comparisons across an entire country. This is achieved through the novel concept of natural cities (c.f., Section 2 for more details or Figure 2 for an illustration). Second, unlike previous studies using census data and satellite imagery, we adopt massive and freely available OSM data for measuring urban sprawl. Consequently, the resulting data can be of value to various urban studies. Following the spirit of the current research at the frontier of data-intensive computing - the fourth paradigm in scientific discovery (Hey et al. 2009), we will release all codes and data sources developed in the study. A third contribution is with the division between sprawling and normal, and between compact and normal. This division is automatically determined according to the head/tail division rule. The head/tail division rule refers to the regularity that *given a variable V, if its values v follow a heavy tailed distribution, then the mean (m) of the values can divide all the values into two parts: a high percentage in the tail, and a low percentage in the head* (Jiang and Liu 2010). A heavy tailed distribution refers to one of the distributions of the power law, exponential, lognormal, and their degenerated versions such as stretched exponential and power law with a cutoff (Clauset et al. 2009). A heavy tailed distribution differs fundamentally from a normal distribution which is often considered to have a thin tail. A similar term, the long tail distribution, is used to refer to a power law like distribution only (Anderson 2006) in discussing a new business model in the internet age.

The remainder of this paper is structured as follows. Section 2 introduces the OSM data and two methods of deriving natural cities from massive street nodes and street blocks. In Section 3, we develop a new method of measuring urban sprawl based on the correlation between natural city sizes and street nodes. We further verify the method in Section 4 by conducting a case study using US census data of population and urban areas, and find that the categorization of sprawling and compact cities matches pretty well with the results in the literature. A second case study is reported in Section 5, where we apply the same method to three European countries: France, Germany, and the United Kingdom. Finally, the paper draws a conclusion and points to the future work.

**2. Natural cities extracted from the massive OSM data**
We intend to abandon any census or statistical data for studying or measuring urban sprawl, since they are not universally or publicly available. Instead we adopt the freely available OSM data. The data are particularly complete and have a large coverage for developed countries. In this section, we will introduce the OSM data and illustrate how natural cities are defined and derived from clustered street nodes and street blocks.

**2.1 The OSM data**
Inspired by the success of Wikipedia, Steve Coast set to map streets of London using GPS traces based on individuals' voluntarily contributions in year 2004. In year 2006, together with some other volunteers, Coast established the OSM foundation, a non-profit organization aimed to create free street map of the entire world using various free data sources. As one can imagine, the mapping is tedious and time consuming even with the active participation of volunteers. In December 2006, Yahoo! granted the right of their aerial photography to the OSM community, so that individuals can digitize streets from the imagery. In April 2007, a Dutch data collection company Automotive Navigation Data donated a complete road dataset of the Netherlands. For the first time, OSM has a full coverage of an entire country. There have been other countries and regions that follow the step of donating their data to OSM. In October 2007, the OSM community completed the transformation of the US Census TIGER road data that are publicly available. A few hundred of gigabytes of OSM data have been collected, although there is a huge inequity in terms of data coverage and data quality. For example, Europe and North America have a very good coverage as well as good data quality, but not the other parts due to restricted national policies on geospatial data. The registered OSM contributors or users have been in a steady growth. In March 2009, the users surpassed over 100,000. Within less than one year in January 2010, this number increased up to 200,000, and now the end of 2010, over 300,000 registered users.

The emerging OSM data provide an important means for GIS and urban related studies. They are massive and owned by no one. Goodchild (2007) has called on GIS community to study VGI in general and OSM in particular. By writing of this paper, we noted several OSM books available in Amazon (e.g., Bennett 2010, Ramm et al. 2010). For the past two years, the researchers have been enjoying playing with the gigabytes of geospatial data. Using the OSM data, we developed an intelligent route service, namely FromToMap (http://fromtomap.org/), aimed to derive the fewest-turn route between two locations. In the web service, we build up a huge graph involving 10 million nodes and 17 million links for the entire Europe street networks. The key technology behind the service is reported in a recent paper (Jiang and Liu 2011).

**2.2 Two methods of deriving natural cities from street nodes and street blocks**
Both street nodes and street blocks are basic units for deriving natural cities. Thus there are two kinds of natural cities: one formed from street nodes and another from street blocks. It should be noted that the computing processes of deriving the natural cities are very intensive, although we have made every effort in optimizing the algorithms. For example, we did not use any commercial GIS software tools; it is impossible for them (e.g. ArcGIS) to handle such massive data. It is also impossible for an ordinary desktop to achieve such a data-intensive computing; instead, we relied on a 64 bits machine (with 4 cores CPU, 48 GB memory, and 1TB hard disk). The method of deriving natural cities using street blocks is considered to be far more time consuming than the one using street nodes. Therefore, for the USA natural cities we derived them from street nodes, while for the three European cases, we derived the natural cities from street blocks.

Before compute the natural cities, we first have to extract the street nodes and street blocks from the massive OSM data. Street nodes refer to both street intersections and street ends, and each street block forms a minimum ring or cycle (*c.f.,* Figure 1 for an illustration). To extract street nodes from a street network, we have to run a line-line intersection operation, something like building up topological relationships of street segments or arcs in most commercial GIS software. At a country level, street nodes have a high density in cities and a low density in the countryside. At a city level, street nodes have a high density in the center and a low density in the periphery. This kind of density distribution matches very well to that of population. We will further assess the correlation between population and street nodes in the following section. Once the topological relationships are built up, we can rely on them to extract street blocks. This is achieved through some traversal processes among the directed arcs. To avoid too technical details about the computing process, we refer the reader to Jiang and Liu (2010) for a complete description.

Figure 1 illustrates the two methods of deriving natural cities: one from street nodes and another from street blocks. The first method is based on the nearest neighbor analysis of street nodes, and clusters those nodes nearby to form individual natural cities. It starts from any node, and draws a circle around it to see which nodes are within the circle. This process goes recursively until no other nodes within a certain radius. The clustering process method is adopted from a city clustering algorithm that was developed by Rozenfeld et al. (2008) for clustering population centers into cities or urban areas. It should be noted that the population centers are defined at an aggregate level rather than at an individual level as street nodes. We adopt the clustering process, and apply it to clustering street nodes instead of population centers. Eventually, we generate several million natural cities for the entire USA, and the city sizes range from 8 million nodes down to a single node (Jiang and Jia 2010). This is the very dataset that will be adopted for measuring urban sprawl in the first case study in the paper.

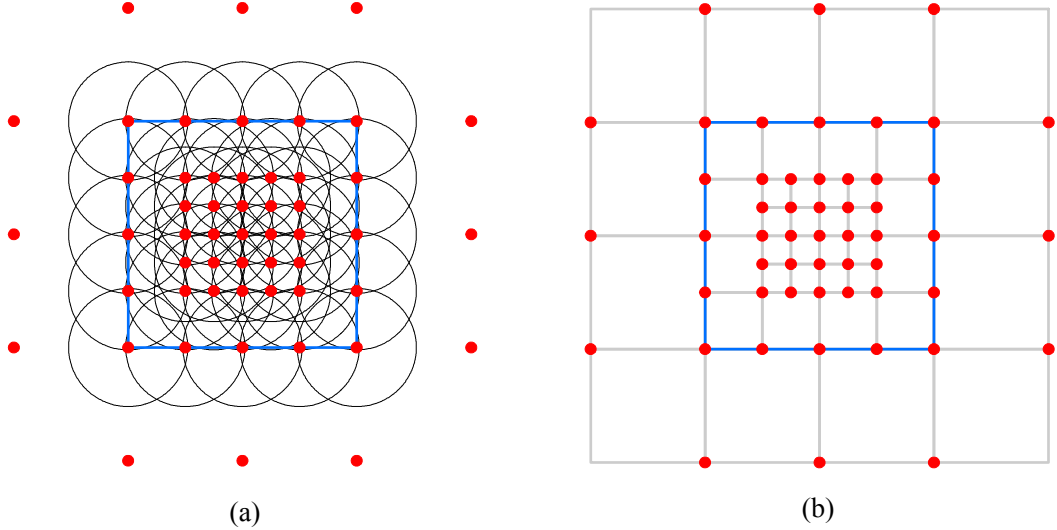

(a)                      (b)

Figure 1: (Color online) Illustration of the two methods of deriving natural cities from street nodes (a) and from street blocks (b) (Note: red dots = street nodes, gray grids = street blocks, black circles = clustering radius, blue box = natural city boundary)

The second method relies on clustering street blocks to derive natural cities. We found that the sizes of street blocks exhibit a lognormal distribution, implying that there are far more smaller (than the mean) blocks than larger (than the mean) ones. According to the head/tail division rule mentioned earlier in the text, all the blocks can be put into two categories: the smaller and the larger. We further group or cluster the smaller blocks into urban agglomerations or natural cities by considering the spatial autocorrelation effect, i.e., group those smaller blocks whose neighbors are also smaller. Note that the underlying idea of the clustering process is similar to clustering street nodes, but we adopt a different principle of defining natural cities. Instead of taking all groups as natural cities, we define only those in the head (of the long trail distribution of groups) as natural cities. Thus there are only a few thousand natural cities for each of the three European countries: France, Germany, and the UK (Jiang and Liu 2010). The very dataset will be adopted for the second case study. It should be noted that the new principle makes better sense, since not all urban agglomerations or human settlements are qualified to be natural cities. In this respect, the second method of defining natural cities seems more natural in terms of determining the head part (or natural cities) according to the head/tail division rule. It should be noted that natural cities and real cities are not completely equivalent, since the latter is legally or administratively determined. However, most natural cities can find their counterparts of real cities. The deviation between natural cities and their corresponding real cities warrants a detailed investigation and comparison.

**3. New method of measuring urban sprawl**
We adopt a simple criterion to determine whether or not a city is sprawling, i.e., a city is considered to be sprawling if urban expansion is faster than population growth. To the contrary, a city is considered to be compact or normal if urban expansion is slower than or equal to population growth. In this paper, we replace the US census urban areas by the defined natural cities and the US census population centers by street nodes for measuring urban sprawl. Figure 2 illustrates two sprawling cities, two compact cities and many normal cities that are closely along the regression line. The gray bandwidth is automatically determined by the measure value of the distances far from the regression line according to the head/tail division rule; refer to the following case studies for more details. The urban expansion for cities along the regression line has the same rate as population growth. In the literature, there are two different views about the relationship between urban expansion (x) and population growth (y). The first is as we illustrated here the linear relationship, i.e., $y = kx$. The second view assumes a nonlinear relationship or a power relationship, i.e., $y = x^\alpha$, where $\alpha < 1$. The second view is also known as the economy of scale, implying that the larger the cities, the less infrastructure per capita, such as street networks, gas stations, and water pipelines (e.g., Bettencourt et al., 2007). The second is built on the theory of allometry initially developed from biology on the study on the growth of part of an organism in relation to that of the entire organism (West et al. 1999). In what follows in this section, we will use the population and urban areas from the US census to investigate the relationships.

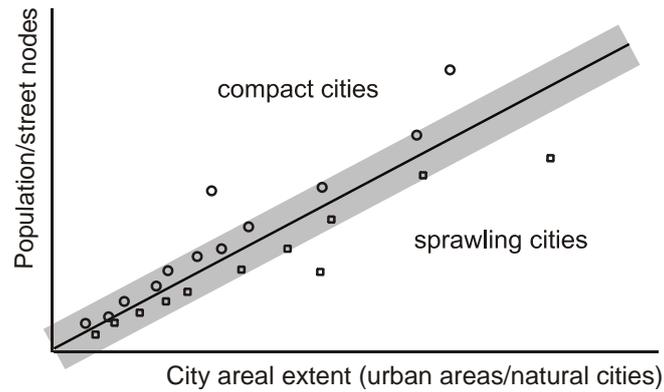

Figure 2: Illustration for differentiating between sprawling and compact cities (Note: cities along the regression line are considered to be normal)

Urban areas are one of the formally defined geographic areas in the US census 2000 data. Urbanized areas (with population > 50,000) and other urban entities (with population between 2,500 and 49,999) are qualified for being urban areas. In other words, the areas or places with less than 2,500 persons are excluded from being urban areas. We downloaded the data from US census (2000a). There are initially 11,880 urban areas, but some of them have the same names or have no name at all. For those without a name, we merged them into large adjacent urban areas. For those with the same name, we also merged them into one unit. Eventually, we obtained 3,638 urban areas for our investigation; refer to Figure 3 for an illustration, which is better viewed in color online rather than in the black/white printed version.

Population data contain population information at the level of census tracts for individual population centers. There are a total of 65,997 population centers, each ranging from 1 to 36,146 people. The data were downloaded from US census (2000b) (excluding 307 invalid records because *x, y* and *pop* are all set to zero). Each entry of the data is uniquely identified by 11 digits, e.g., for the first entry "01001020100, 1921, +32.47507, -86.486814". The first 2 digits correspond to the state, the next 3 to the county within the state and the rest to the census tract. The first record indicates some state (01), some county (001), and some census tract (020100), with population 1921 located at +32.47507, -86.486814. Overlapping urban areas and the population data, we found that there are many population centers that are not within any urban area. Among the 65,997 population center, only 49,114 of them are within urban areas, accounting for 76% of the entire population.

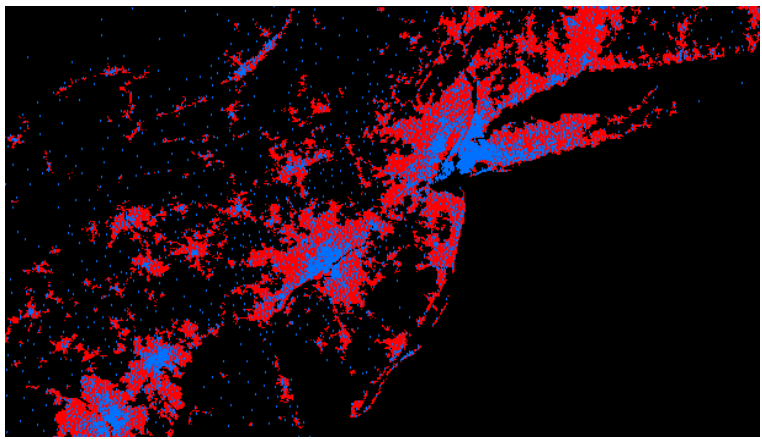

Figure 3: (Color online) The US census (2000a, 2000b) Urban Areas (red patches) and Population Centers (blue points) near New York

We examined the correlation between urban areas (i.e., areal extent) and population, and found that the R square for the linear relationship is around 0.83. On the other hand, if we took the log-log plot, i.e. $\log(y) = \alpha \log(x)$, we found that the R square for the power relationship is 0.85, and $\alpha = 0.95$. The two R square values are almost the same (deviation is 0.02), and α is very close to 1.0. It appears very difficult to differentiate between a linear and nonlinear relationship between urban areas and populations. It is important to point out a major problem of population data. They are population areas aggregated to population centers. Unlike the population data, street nodes are defined at the individual level, it is our intention to replace population by street nodes. Let us assess how population is significantly correlated to street nodes. We take the 3638 urban areas and plot them against street nodes within the urban areas, and find that the correlation coefficient R square is 0.89. This significant correlation has proved that street nodes defined at the individual level can be a proxy for population for measuring urban sprawl.

**4. Case study I: measuring urban sprawl of the natural cities of USA**
We adopt over 30 thousand of the largest natural cities from over 3 million of natural cities derived from the previous study by Jiang and Jia (2010) for our investigation. The smallest natural city contains only 56 street nodes. There are 14 million of street nodes within the 30 thousand of natural cities. The boundaries of natural cities are delineated by imposing a grid of resolution of 500 meters. We plot natural cities against street nodes (Figure 4), and find a significant correlation between two parameters, with an R square up to 0.96.

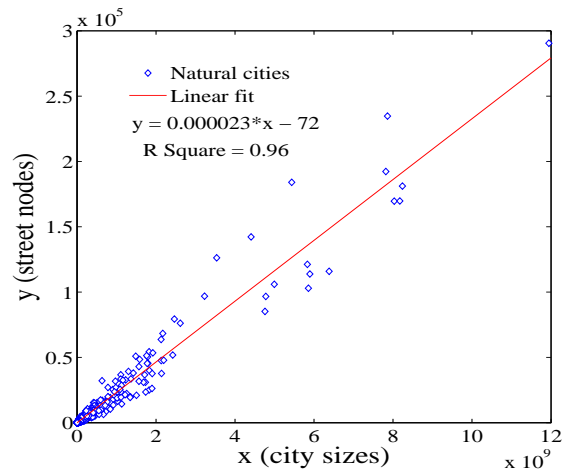

Figure 4: (Color online) Correlation plot between natural cities (physical areas) and street nodes for USA

As we can see, a vast majority of the natural cities are along the regression line, and only a very few natural cities are far from the line. The fact that most dots are clustered around that corner of the plot indicates another important fact that the sizes of the natural cities follow a power law distribution; refer to Jiang and Jia (2010) for more details. We further compute the distance between points (representing natural cities) and the regression line, and find that the distance exhibits a striking lognormal distribution. With respect to Figure 4, we partition all points (representing 30 thousand of the natural cities) into two groups: (1) those above the regression line and (2) those below the line. For each group, we examine whether or not their distances far from the regression line follow a heavy tailed distribution. The plot shown in Figure 5 demonstrates the striking lognormal distribution.

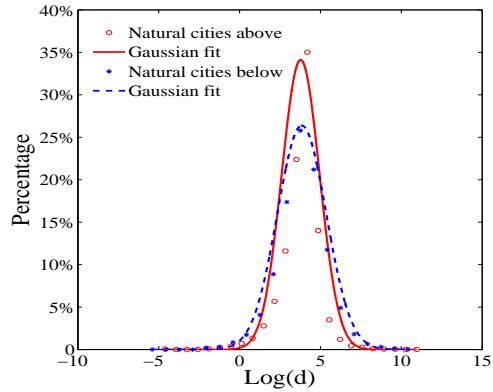

Figure 5: (Color online) Lognormal distribution of the distance between individual points and the regression line in Figure 4

Given the lognormal distributions, we obtain a mean for each group to partition the corresponding natural cities into two. For the first group (above the line), those far from the regression line are considered to be compact cities, and those close to or along the regression line are considered to be normal cities. For the second group (below the line), those far from the regression line are classified as sprawling cities, and those close or along the regression lines as normal cities. The classification between compact and normal (above the line) and between sprawling and normal (below the line) is a direct application of the head/tail rule (Jiang and Liu 2010). The reason why we can make such a classification can be justified as such. If some things exhibit a heavy tailed distribution, the mean value would help to identify abnormal things. In this regard, we can remark that most cities are normal cities, only a few cities are sprawling or compact. Therefore, we can identify the very few cities using the mean value. We choose the top 500 natural cities for visualization using traffic light colors: red for sprawling cities, green for compact, and yellow for normal cities. The result is shown in Figure 6. We can note that most large cities are either sprawling or compact, and only very small cities (nearly invisible) are normal.

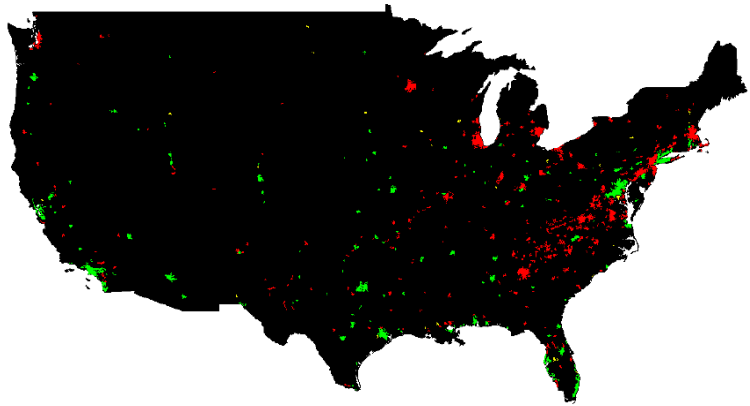

Figure 6: (Color online) Top 500 natural cities are classified into sprawling (red), compact (green) and normal (yellow, invisible due to smaller sizes) according to natural cities and street nodes correlation

To verify the above result, we plot urban areas against street nodes, and identify all sprawling, compact and normal cities as shown in Figure 7. It is not hard to note that the two patterns shown in Figure 6 and Figure 7 are very consistent. The resulting classification is also very consistent with the existing literature (Sutton 2003). We further choose the top 25 cities for a comparison and find only three inconsistent cases.

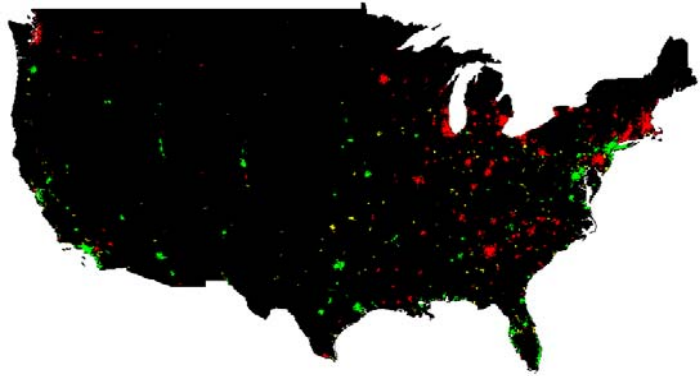

Figure 7: (Color online) The top 500 urban areas are classified into sprawling (red), compact (green) and normal (yellow, invisible due to smaller sizes) according to urban areas and street nodes correlation

Let us take a detailed look at the three inconsistent cases. With respect to Figure 8, natural cities are shown in blue, while urban areas are in red. For the first case (Figure 8a), natural city C is sprawling, while the corresponding urban area (including parts that are overlapped with C and A) is compact. Note that the urban area is far larger than natural city C. This is the first inconsistency. In fact, part A and B together constitute another natural city which is compact. To investigate this inconsistency, we separate A and B, and examine their sprawling or compact level. We find that part A is extremely compact, and part B is slightlysprawling, so A and B together as one is still compact. In the same way, natural city C is slightly sprawling, but A and C together as one is compact due to a major contribution of A. A and C together as one has the same boundary as the urban area. For the second and third cases (Figure 8b and 8c), natural cities are far larger than the corresponding urban areas. That can explain why there is an inconsistency in the classification between compact and sprawling. After the detailed examination of the inconsistent cases, we have seen that the inconsistencies are mainly due to the inconsistent boundaries of natural cities and urban areas. It proved from another perspective that the approach to measuring urban sprawl relying on street nodes and natural cities is reliable and valid.

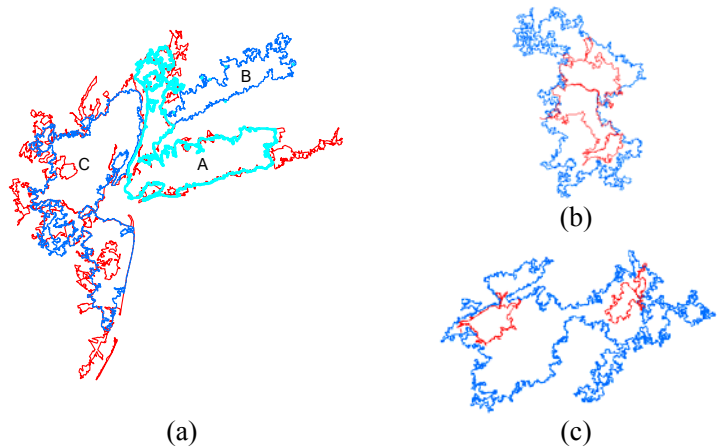

(a)     (b)     (c)

Figure 8: (Color online) Three inconsistent cases between urban areas in red and natural cities in blue: (a) New York region, (b) Richmond, VA, and (c) Roanoke and Lynchburg, VA.

What if we adopt the US census data of population and urban areas for a similar investigation? The result is again similar to what we have illustrated above (c.f., Figure 9). For the top 25 natural cities or urban areas, we find only three inconsistant cases. Firstly, Chicago changes from early sprawl to compact. This change can be blamed on the use of population data, since the exisitng literature (e.g., Sutton 2003) supports our result that Chicago is a sprawling city. The second case changes from early compact to normal. This is indeed a little change. More importantly, the natural city has no corresponding urban area, which makes little sense to the comparison. The third case changes from early compact to sprawling, a dramatic change indeed. However, we note that the boundary is increased dramatically, which can justify the change.

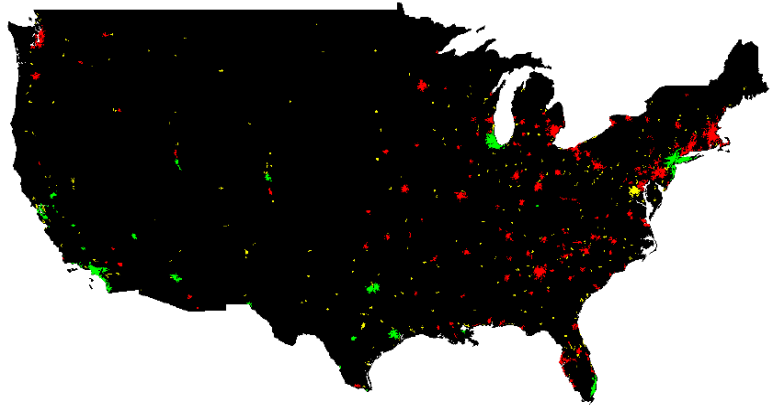

Figure 9: (Color online) Top 500 urban areas are classified into sprawling (red), compact (green) and normal (yellow, many invisible due to smaller sizes) according to urban areas and population correlation

**5. Case study II: measuring urban sprawl of three European countries**
Given the promising result from the first case study, we further apply the approach to three European countries: France, Germany and the United Kingdom. With the second case study, we adopt natural cities that are delineated from street networks rather than through clustering street nodes. The second way of defining natural cities is guided by the head/tail division rule (Jiang and Liu 2010), an interesting regularity that can characterize many natural and societal phenomena in terms of the inbuilt balance between the head and the tail of a heavy tailed distribution. Under the revised definition, cities are considered to be at the head part of a long tail distribution of human settlements. The revised definition of natural cities makes better sense since the early definition of natural cities tend to include all human settlements down to the smallest with only a single person.

We first plot the natural cities against street nodes, and subsequently note a significant correlation between the two parameters as shown in Figure 10. This is the base for measuring urban sprawl. Next, we again find that the distances between the points (representing individual natural cities) to the regression line follow a striking lognormal distribution (Figure 11), although the mean and standard deviation for the parameter log(x) vary from one country to another. Following the same procedure in the first case study, we classify all natural cities into three categories as shown in Figure 11.

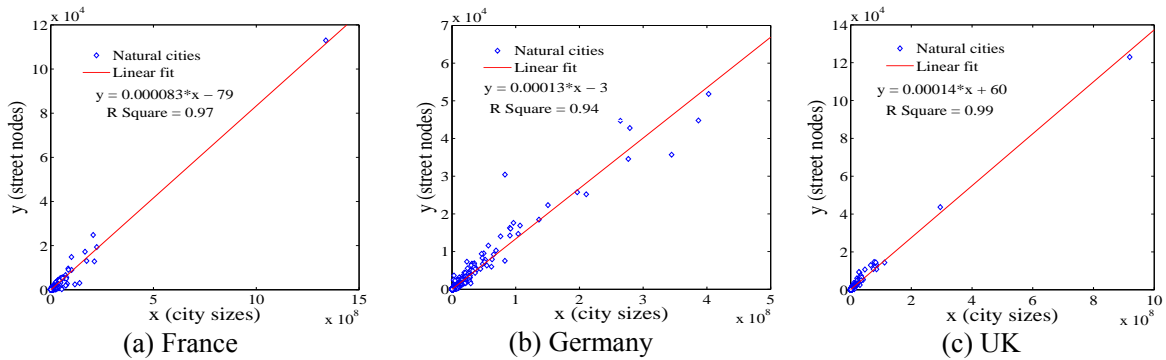

(a) France (b) Germany (c) UK

Figure 10: (Color online) Correlation plot between natural cities (physical areas) and street nodes for the three European countries

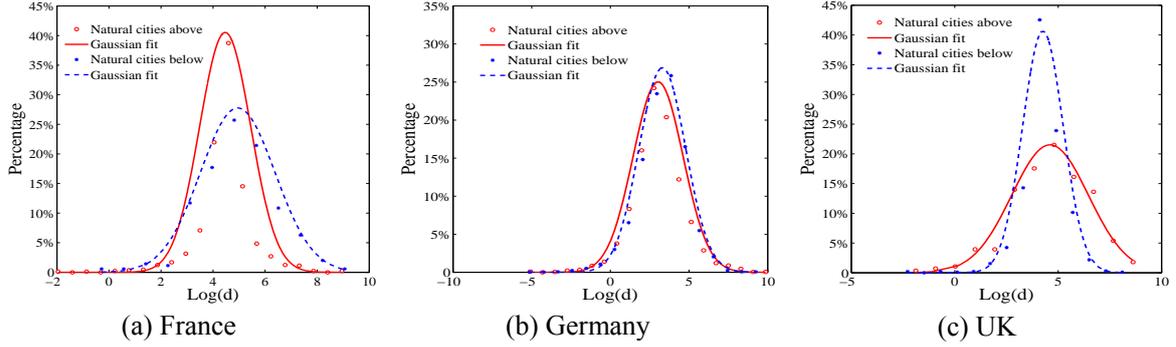

| (a) France | (b) Germany | (c) UK |

Figure 11: (Color online) Lognormal distribution of the distance between individual points and the regression line in Figure 9

To this point, we have classified all the natural cities into three categories for the three European countries; the statistics is presented in Table 1. As mentioned above, this result is based on some uniform standard that is comparable across an entire country. However, how this result about sprawling or compact cities is related to a situation in reality warrants further comparison or verification due to a simple fact that the boundaries of natural cities and real cities could be different. As illustrated above for the US case study, the relationship between natural cities and real cities is not just one-to-one, but one-to-two, or one-to-many. Also, there is no such a guarantee that the boundaries of natural cities completely match those of real cities. However, the above result provides a benchmark dataset for cross comparisons. It also provides an effective standard to determine whether or not a city is sprawling given the naturally determined city boundaries. We believe the overall result matches pretty well with the general perception about city sprawl or compact. For some cases where real city boundaries deviate too much from natural cities, we need to do the same in-depth examination as we did above for the three inconsistent cases with the USA.

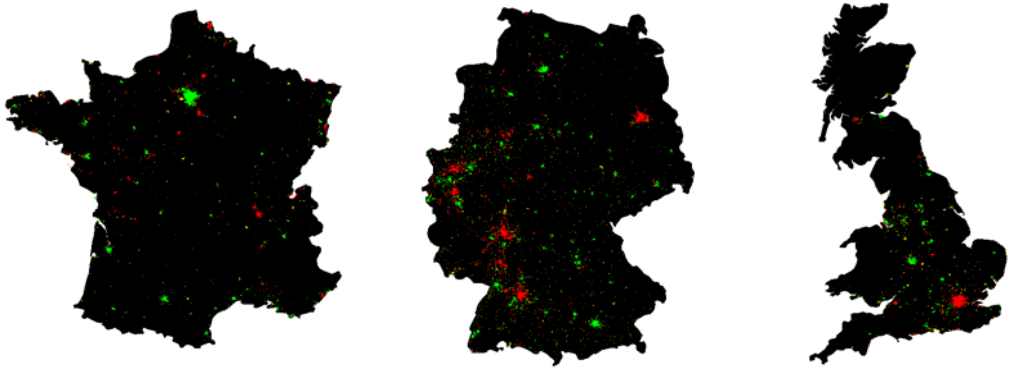

Figure 12: (Color online) All natural cities are classified into sprawling (red), compact (green) and normal (yellow, invisible due to smaller sizes) for the three countries

Table 1: Statistics about sprawling and compact cities in the three European countries

| Country | Natural cities | Street nodes in cities | Sprawl cities | | Compact cities | |
|---|---|---|---|---|---|---|
| | | | Number | Percentage | Number | Percentage |
| France | 1,238 | 606,324 | 79 | 6% | 149 | 12% |
| Germany | 5,124 | 1,619,228 | 687 | 13% | 302 | 6% |
| UK | 1,245 | 817,029 | 270 | 22% | 73 | 5% |

Compared to previous studies relying on population and urban areas or urban boundaries auto-detected from remote sensing imagery, our approach to measuring urban sprawl has some advantages. First, street nodes are defined at the individual level rather than as an aggregate level as with populations. Thus, they can be a good proxy for populations defined at the individual level. Second, unlike urban areas imposed from the top down through legal or administrative means, natural city boundaries are formed from the bottom up and they are naturally and automatically delineated. In this regard, natural cities provide an alternative to the city boundaries auto-detected from remote sensing imagery. However, our approach is not without any problems. The OSM data for the four developed countries have a very good coverage, and we can trust the data quality for such kind of analysis. Yet for many developing countries, the OSM data is far from complete and reliable. In addition, the OSM data do not allow us to measure urban sprawl from the temporal dimension. This is probably one critical weakness when compared to remote sensing imagery for delineating city boundaries.

**6. Conclusion**
This paper developed a novel approach to measuring urban sprawl using street nodes and natural cities. We abandoned the use of population and city boundaries for the measurement, since these data are defined at an aggregate level and are criticized for being subjective or arbitrary due to legal and administrative factors. Instead we adopted universally available geospatial data voluntarily contributed by individuals, and supported by the Web 2.0 technologies. This is a bottom up approach. The natural city boundaries were objectively and effectively derived. The street nodes and natural cities replace population and real cities for the measurement of urban sprawl. The developed approach was verified by the US census data of population and urban areas. We found very consistent results in terms of classifying sprawling or compact cities. We further apply the approach to three European countries, and categorized all natural cities into three classes: sprawling, compact and normal.

Together with the methodological development, we provided valuable datasets about the classification of natural cities for the four developed countries across the Atlantic. All the datasets, including street nodes, natural cities, and their classification, will be released freely for further studies. Interested researchers are encouraged to contact us for access to the data. We hope the data can be a benchmark for various urban studies in the future. Following a key spirit of data-intensive computing, we will also release the source codes developed in the study. As for the future work, there are many things to be done. Our current investigations focus mainly at a macro or global level. Further studies should can be done at a micro and detailed level in terms of measuring urban sprawl, in terms of how natural cities match the US census' urban areas, and in terms of cross comparison of urban sprawl between USA and Europe.

**Acknowledgement**
The data used in this study were taken from the OSM databases. We will make all the related data freely available for academic purposes – "freely received and freely give". Interested individuals are encouraged to contact us for access to the data. We thank Xintao Liu for his help with Figure 1.